\documentclass[useAMS,usenatbib]{mn2e}
\newcommand{\LA}{\mbox{\raisebox{-0.6ex}{$\stackrel{\textstyle<}{\sim}$}}}

\usepackage{graphicx}
\title{The Kinetically Dominated Quasar 3C~418}
\author[Brian Punsly and Preeti Kharb]{Brian Punsly and Preeti Kharb\\
1415 Granvia Altamira, Palos Verdes Estates CA, USA
90274 and ICRANet, Piazza della Repubblica 10 Pescara 65100, Italy\\
\\ National Centre for Radio Astrophysics, Tata Institute of
Fundamental Research, Post Bag 3, Ganeshkhind, Pune 411007, India\\
E-mail: brian.punsly@cox.net}
\begin{document}
\maketitle \label{firstpage}
\begin{abstract}The existence of quasars that are kinetically dominated, where the
jet kinetic luminosity, $Q$, is larger than the total (IR to X-ray)
thermal luminosity of the accretion flow, $L_{\rm{bol}}$, provides a
strong constraint on the fundamental physics of relativistic jet
formation. Since quasars have high values of $L_{\rm{bol}}$ by
definition, only $\sim 10$ kinetically dominated quasars (with
$\overline{Q}/L_{\rm{bol}}>1$) have been found, where $\overline{Q}$
is the long term time averaged jet power. We use low frequency (151
MHz$-$1.66 GHz) observations of the quasar 3C\,418 to determine
$\overline{Q}\approx 5.5 \pm 1.3 \times 10^{46} \rm{ergs~s^{-1}}$.
Analysis of the rest frame ultraviolet spectrum indicates that this
equates to $0.57 \pm 0.28$ times the Eddington luminosity of the
central supermassive black hole and $\overline{Q}/L_{\rm{bol}}
\approx 4.8 \pm 3.1$, making 3C\,418 one of the most kinetically
dominated quasars found to date. It is shown that this maximal
$\overline{Q}/L_{\rm{bol}}$ is consistent with models of
magnetically arrested accretion of jet production in which the jet
production reproduces the observed trend of a decrement in the
extreme ultraviolet continuum as the jet power increases. This
maximal condition corresponds to an almost complete saturation of
the inner accretion flow with vertical large scale magnetic flux
(maximum saturation).
\end{abstract}
\section{Introduction}
The relationships among jet production, accretion state and central
supermassive black hole parameters are not well understood. While
radio quiet quasars (RQQs) make the majority of quasars, only
$\approx10\%$ of optically selected quasars are radio loud quasars
(RLQs), and only about $2\%$ of them have powerful extended radio
lobes \citep{dev06}. Relativistic jets occur in low luminosity
active galactic nuclei (LLAGN) and in powerful RLQs. Curiously, the
long term time averaged jet power normalized to the Eddington
luminosity, $Q_{\rm{Edd}}$, of the LLAGN, M87, is four orders of
magnitude less than the value for the quasar 3C\,418 derived in this
paper. A fundamental issue is therefore raised: is it possible that
the same jet launching mechanism will span four orders of normalized
jet power? In order to make questions like this well posed, we
continue our search for the most powerful jets. In kinetically
dominated quasars, the jet kinetic luminosity, $Q$, is larger than
the total (IR to X-ray) thermal luminosity of the accretion flow,
$L_{\rm{bol}}$ \citep{pun07}. This extreme condition is valuable
observational data for constraining jet launching models. The
kinetically dominated quasar condition is very rare since the
definition of quasars requires  a large optical/UV luminosity,
$M_{V}<-23$ which equates to $L_{\rm{bol}}> 3 \times 10^{45}
\rm{ergs~s^{-1}}$ \citep{pun06}. Yet, the maximum jet powers are
only $Q \sim 10^{47} \rm{ergs~s^{-1}}$ \citep{wil99}. The quasar,
3C\,418, is one of the the most luminous quasars in the Third
Cambridge (3C) Catalog of Radio Sources with a 178~MHz flux density
of 13.1~Jy at a redshift of z = 1.686 \citep{smi80}. In spite of
this, there are few pointed observations of this source, primarily
due to the low Galactic declination of $\sim 6^{\circ}$ and the
associated large Galactic extinction. Since it is one of the most
luminous low frequency radio sources in the known Universe, this an
optimal candidate for a kinetically dominated quasar and is the
subject of the in depth study presented here.

A shortcoming of this analysis is that in order for $L_{\rm{bol}}$
and $Q$ to be estimated contemporaneously necessitates that $Q(t)$
be derived from models of parsec scale radio jets, however the
potential large uncertainty due to a poorly constrained Doppler
enhancement is the topic of debate \citep{pun05,pun06,ghi15}. By
contrast, the time averaged jet power $\overline{Q}$ can be
estimated more accurately from the isotropic properties of the
extended radio lobe emission \citep{wil99}. Unfortunately, the
$\overline{Q}$ estimate is not contemporaneous with the
$L_{\rm{bol}}$ data, so one cannot say if the sources presently
satisfy or ever satisfied $R(t)=Q(t)/L_{\rm{bol}}>1$. In spite of
this obstacle, we argued in \citet{pun07} that if $\overline{Q}$
exceeds the Eddington luminosity of the central supermassive black
hole then it is very likely that $R(t)=Q(t)>1$ at some instance of
the quasar lifetime since $L_{\rm{bol}}/L_{\rm{Edd}}>1$ states are
very rare for quasars \citep{gan07}. Thus motivated, we make an
argument that $\overline{Q}/L_{\rm{Edd}}\sim 1$ for 3C\,418.
\par In Section 2, we review estimation techniques for
$\overline{Q}$. These methods rely on the low frequency spectrum of
the radio lobes. In Section 3, we present previously unpublished
observations in order to find the 151~MHz and 330~MHz flux
densities. There is a large low frequency excess over what is
expected from the radio core and jet. We use this information in
order to estimate the lobe flux and $\overline{Q}$. In Section 4, we
de-redden the ultraviolet spectrum to estimate $L_{\rm{bol}}$ and
use the MgII line width to estimate the central black hole mass. The
data are synthesized in Section 5. In this paper, we adopt the following
cosmological parameters: $H_{0}=70$~km~s$^{-1}$~Mpc$^{-1}$, $\Omega_{\Lambda}=0.7$
and $\Omega_{m}=0.3$.  We define the radio spectral index, $\alpha$,
as $F_{\nu}\propto\nu^{-\alpha}$.

\section{Estimating Long Term Time Averaged Jet Power} The more
information that is known about the radio lobes such as the radio
spectral index across the lobe and high resolution X-ray contours,
the more sophisticated and presumably more accurate the estimate of
$\overline{Q}$ \citep{mcn11}. Unfortunately, such detailed
information does not exist for most radio sources, including
3C\,418, and a more expedient method is required. Such a method that
allows one to convert 151~MHz flux densities, $F_{151}$ (measured in
Jy), into estimates of $\overline{Q}$ (measured in ergs/s), was
developed in \citet{wil99}. The result is captured by the formula
derived in \cite{pun05}:
\begin{eqnarray}
 && \overline{Q} \approx 1.1\times
10^{45}\left[X^{1+\alpha}Z^{2}F_{151}\right]^{\frac{6}{7}}(\textbf{f/15})^{\frac{3}{2}}\mathrm{ergs~s^{-1}}\;,\\
&& Z \equiv 3.31-(3.65)\times\nonumber \\
&&\left[X^{4}-0.203X^{3}+0.749X^{2}
+0.444X+0.205\right]^{-0.125}
\end{eqnarray}
where $X\equiv 1+z$, and $F_{151}$ is the total optically thin flux
density from the lobes. The formula is most accurate for
large relaxed classical double radio sources. Due to Doppler
boosting on kpc scales, core dominated sources with a very bright
one sided jet must be treated with care \citep{pun05}. 3C\,418 is
dominated by a flat spectrum core and a one-sided jet in the high
resolution 4.86 GHz images with the Very Large Array (VLA)
\citep{ode88}. This defines our primary task of separating the
optically thin lobe emission from the strong core/lobe feature. The
calculation of the jet kinetic luminosity in Equation (1)
incorporates deviations from the overly simplified minimum energy
estimates into a multiplicative factor, \textbf{f}, that represents
departures from minimum energy, geometric effects, filling factors,
protonic contributions and low frequency cutoff \citep{wil99}. The
quantity, \textbf{f}, was further determined to most likely  be in
the range of 10 to 20, hence the fiducial value of 15 in Equation
(1) \citep{blu00}.
\par Alternatively, one can also use the independently derived
isotropic estimator in which the lobe energy is primarily inertial
(i.e., thermal, turbulent and kinetic energy) in form \citep{pun05}
\begin{eqnarray}
&&\overline{Q}\approx
5.7\times10^{44}(1+z)^{1+\alpha}Z^{2}F_{151}\,\mathrm{ergs~s^{-1}} \;.
\end{eqnarray}

\begin{table}
\caption{Component Flux Densities} {\footnotesize
\begin{tabular}{ccccc}
 \hline
Component &   Frequency &  Flux  & Flux  &  Comments  \\
          &      &  Density & Density &    \\
       &   (GHz) &  Peak (Jy) & Integral (Jy)  &  \\
\hline
C0 & 14.94 &  $2.640 \pm 0.132$ & N/A &  1   \\
C0 & 4.86 &  $2.800 \pm 0.140$ & N/A &  1   \\
C0 & 1.66 &  $4.597 \pm 0.230$ & $4.687 \pm 0.234$ &  2   \\
C1 & 1.66 &  $0.351 \pm 0.018$ & $0.439 \pm 0.022$ &  2   \\
C0 + C1 & 1.66 &  $4.597 \pm 0.230$ & $5.264 \pm 0.256$ &  2   \\
C2 + L2 & 1.66 &  $0.276 \pm 0.014$ & $0.358 \pm 0.018$ &  3   \\
C3 + L3 & 1.66 &  $0.085 \pm 0.004$ & $0.107 \pm 0.005$ &  3   \\
A1 & 0.33 &  $7.483 \pm 0.374$ & $7.592 \pm 0.385$ &  4   \\
A2 & 0.33 &  $1.708 \pm 0.086$ & $2.258 \pm 0.113$ &  5   \\
\hline
Core + Jet & 1.66 &  N/A & $5.486 \pm 0.549$ &  6   \\
Core + Jet & 0.33. &  N/A & $7.483 \pm 0.374$ &  7   \\
Lobes & 1.66 &  N/A & $0.327 \pm 0.033$ &  8   \\
Lobes & 0.33 &  N/A & $2.466 \pm 0.247$ &  9   \\
\hline
\end{tabular}}
\footnotesize{1. Jet and lobe flux appears to be
over-resolved in O'Dea et. al. (1988). 2. Resolution insufficient to remove the jet
contribution from the core. 3. Comparing to the 5 GHz images, this
feature appears to have a very steep spectrum. 4. Beam size too large to
resolve core flux from jet and lobes. 5. Most of the lobe flux
density is in this component. 6. Estimated as integral flux of C0 +
C1 and one half the peak flux of C2 + L2. 7. Peak flux density of A1.
8. Contains three pieces: integral flux of C3 + L3, integral$-$peak
flux C2+L2 and and one half the peak flux of C2 + L2. 9. A2 plus the
integral$-$peak flux density of A1 (everything except peak flux of A1).}
\end{table}

Equation (3) generally estimates $\overline{Q}$ lower than
Equation(1). Consequently, we use Equation (1) with
$\mathrm{\textbf{f}} = 20$ as the maximum upper bound on
$\overline{Q}$ and Equation (3) is the lower bound $\overline{Q}$ in
the following.

\begin{figure}
\includegraphics[width=50 mm]{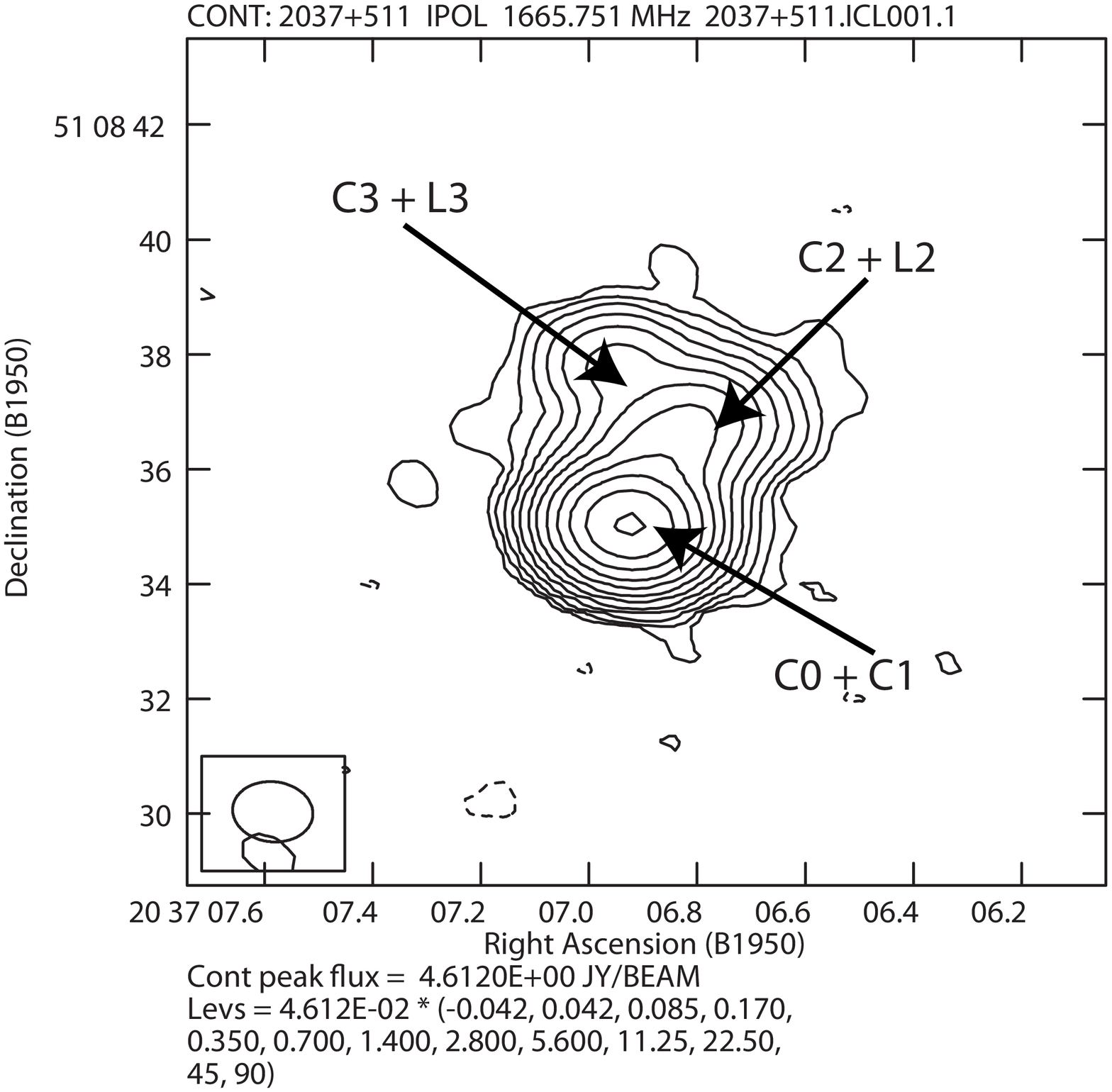}
\includegraphics[width=50 mm]{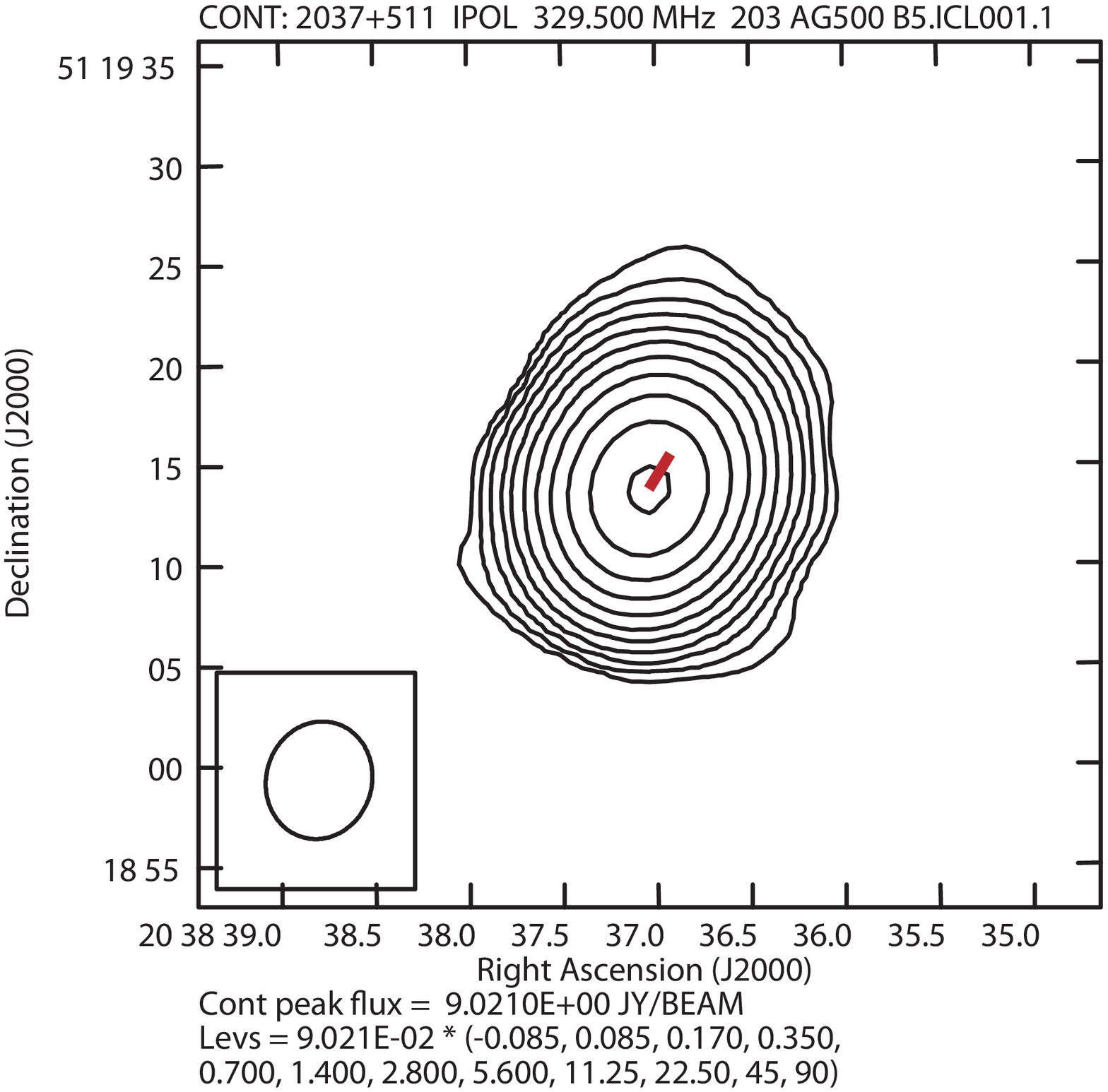}
 \caption{The components in Table 1 are identified in this 1.66 GHz VLA A-array image (top panel).
 The 330 MHz A-array image (bottom panel) is
very slightly resolved in the North-South direction. The red line
indicates the jet from C0 to C2.}
    \end{figure}

\section{Low Frequency Radio Observations}
In this section, we discuss about the archival 151~MHz Giant
Metrewave Radio Telescope (GMRT) data and the newly reduced 330~MHz
VLA data, in combination with previously published 1.66~GHz and
4.86~GHz VLA data, in order to segregate the lobe emission from the
total flux at low frequency. The 151~MHz data is from the
TGSS\footnote{The Tata Institute of Fundamental Research Giant
Metrewave Radio Telescope All Sky Survey} Alternate Data Release 1
\citep{int16}. The beam width is $> 20\arcsec$, much larger than the
size of the source at higher frequency. The data are consistent with
a point source and we choose the peak flux density of
$16.50\pm0.17$~Jy as the best estimate of the total flux of the
compact radio source.

We use the 1.66~GHz VLA A-array radio image from \citet{kha10} in Figure~1 in order to
assess the contribution of the various components. We have made  annotations
based on the high resolution images at 4.86~GHz and 14.94~GHz from \citet{ode88}. The beam in the
top panel of Figure~1 is $1.40\arcsec\times 1.05\arcsec$ at a PA = $83.85\degr$. The core and the
strong inner knot in the jet are partially resolved and are designated as C0+C1. We denote the other
two components as C2+L2 and C3+L3, where C2 and C3 are the compact
components seen at 4.86~GHz in \citet{ode88} and L2 and L3 are the
the diffuse emission associated with these features. The component
C2+L2 is identified with the lobe on the jet side. Due to projection
effects of an almost pole-on orientation, the component C3+L3 could
be the radio lobe on the counter jet side.
\begin{figure}
\includegraphics[width=70 mm]{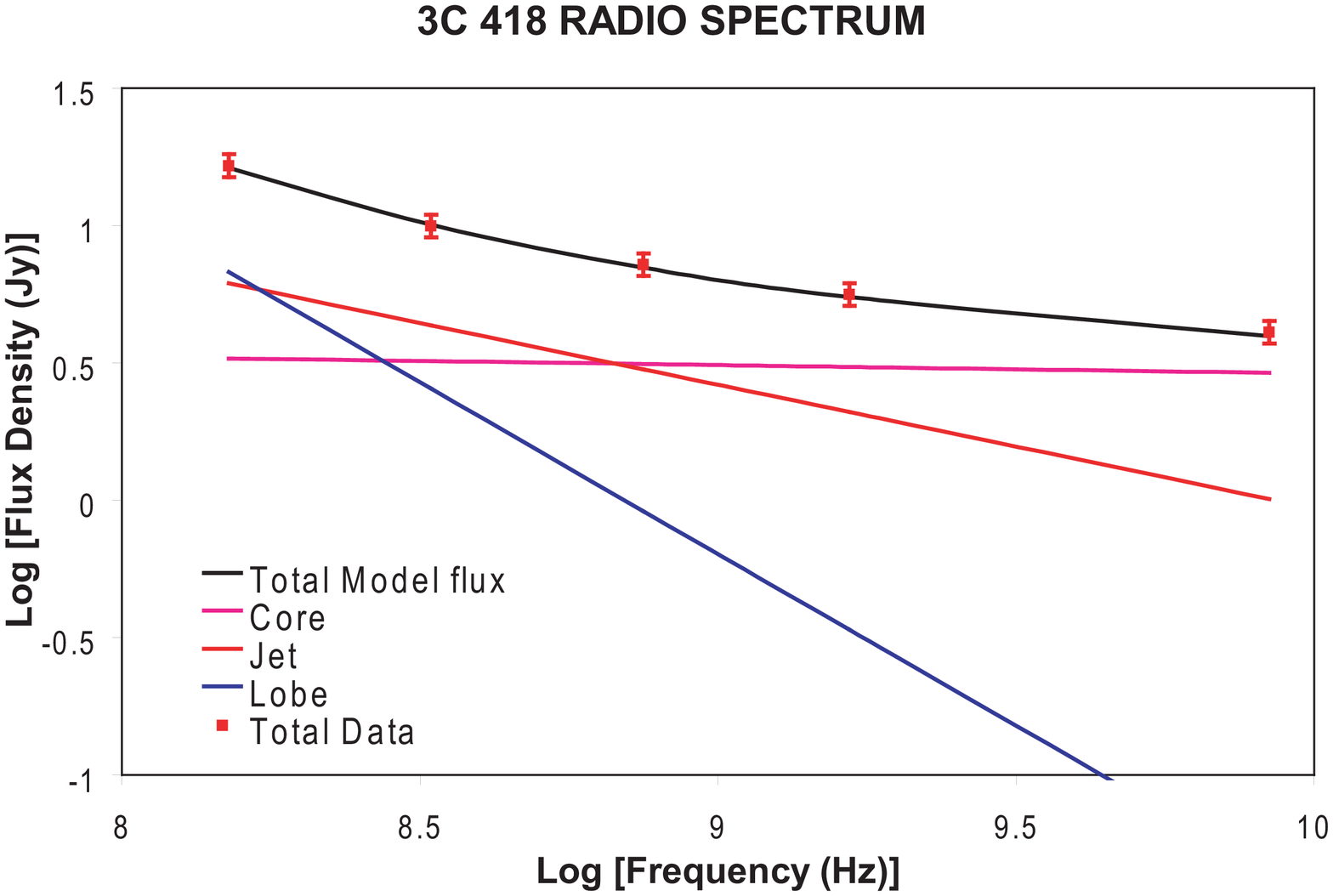}
\includegraphics[width=70 mm]{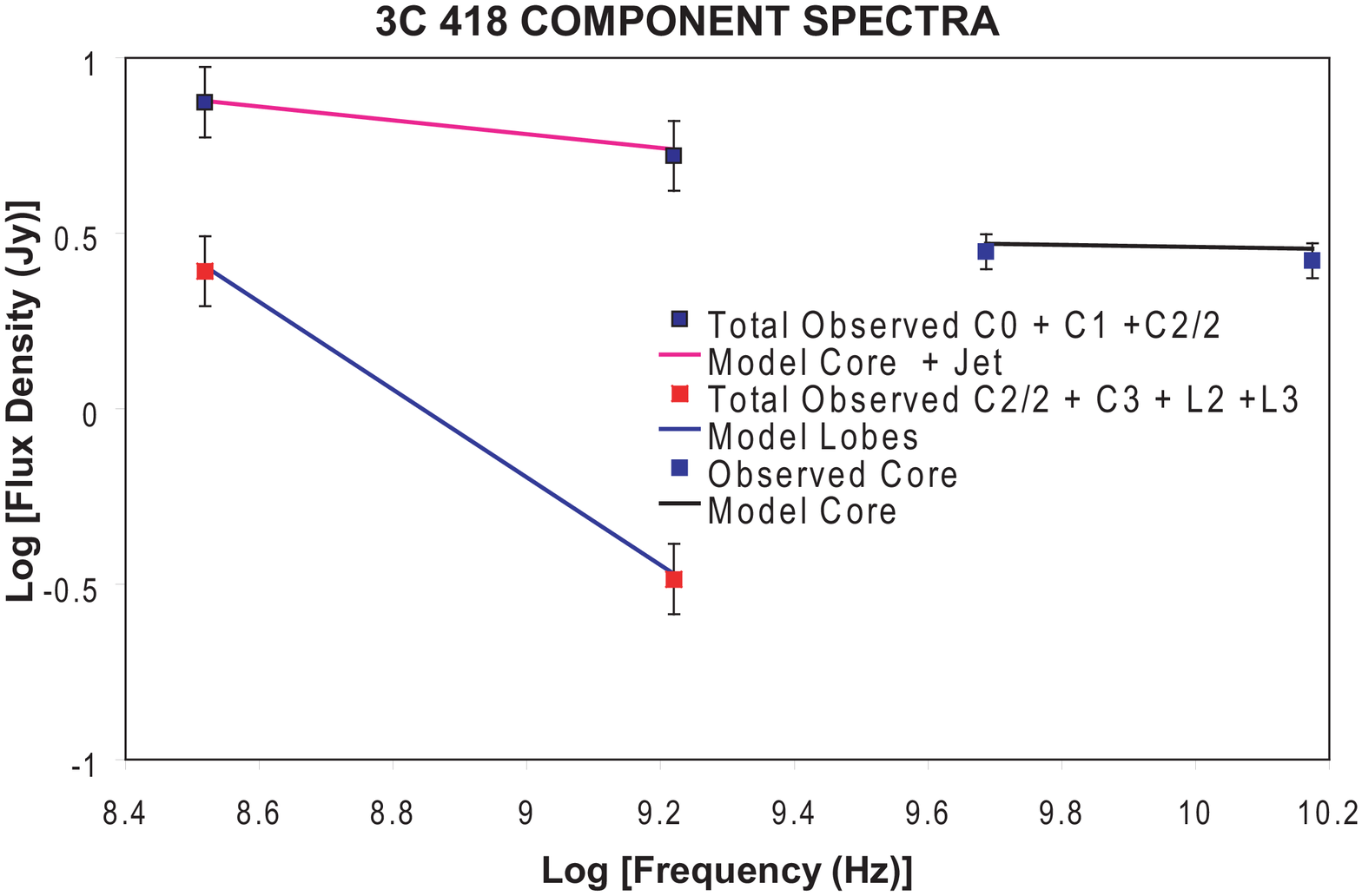}
 \caption{The Model A power law fit to the broadband radio data in Table 2 is
 shown in the top panel. The bottom panel shows the additional constraints of
 simultaneously fitting the components.}
    \end{figure}
The bottom panel of Figure~1 is the 330~MHz VLA A-array image from
the project AG500. These data were reduced using standard procedures
in AIPS\footnote{Astronomical Image Processing System}. The beam is
$5.91\arcsec\times5.27\arcsec$ at a PA = $-16.58\degr$. There is no
large diffuse cloud of extended flux, but there is a slight
elongation in the North-South direction in the image made with this
large beam. The short red line indicates the jet path from C0 to C2
in the top panel at 1.66~GHz.

Table~1 shows the component flux densities based on model fits
to the images. The 1.66~GHz image was modeled with 4 Gaussian
components based on the 4 components seen at 4.86~GHz. It could be
that C0 has varied significantly between the 4.86~GHz and 1.66~GHz
observations, but the two epochs in \citet{ode88} showed minimal (a
few per cent) variability and a very flat spectrum. We note that the
separation between C0 and C1 in \citet{ode88} is much less than the
beam size at 1.66 GHz. Thus, the apparent increased flux in C0 at
1.66 GHz could be just due to the blending of components. The
blending of C0 and C1 at 1.66 GHz makes it hard to model the spectra
of the individual components, hence they are combined in the
following multi-component theoretical modeling of the source. We
tried fitting the 330 MHz image with 1, 2 and 3 Gaussian components
using the JMFIT task in AIPS. The best fit was with 2 components and
the results are in Table 1. We list these as components A1 and A2
since it is not clear {\it a priori} which of the components in the top
panel of Figure~1 are blended by these fits. Our methodology is to
assume a basic blend of components, where A1 is dominated by the
core and jet flux with some lobe flux and A2 is dominated by the
lobe flux. The exact blend that is ultimately chosen is based on
simultaneously fitting the component spectra with the 4-component
theoretical power law models that we now describe.
\par The theoretical model fits are complicated by the fact that the
observationally blended components (OBCs) and the natural
theoretically blended components (TBCs), denoted below the
horizontal line in Table 1, are not exactly the same. Thus, the
model must simultaneously fit the decomposition of TBCs in terms of
OBCs and the power laws of the individual components. We fit the
TBCs separately in addition to the total flux density in Figure 2.
This removes degeneracy that occurs by just fitting the total flux
spectrum. We considered two classes of simple power law models in
Table 2. Model A, in Figure 2, fits the core at high frequency,
where it is resolved from the jet and the TBCs (in the bottom panel)
as well as the total flux density (with additional data from the
NASA Extragalactic Database, NED) in the top panel. Model B assumes that
the core flux density is variable and is not fit separately, thus
the other fits are slightly better. It is not shown because it looks
similar to the model A fits.
\begin{table}
\caption{Model Flux Densities} {\footnotesize
\begin{tabular}{cccc}
 \hline
Component &   Flux Density &  Power Law   \\
Model A   & 151 MHz (Jy)    &  Spectral Index ($\alpha$)   \\
\hline
Core & 3.28 &  0.03   \\
Jet & 6.17 &  0.45   \\
Lobes & 6.76 &  1.25  \\
\hline
Component &   Flux Density &  Power Law   \\
Model B   & 151 MHz (Jy)    &  Spectral Index ($\alpha$)  \\
\hline
Core & 4.08 &  0.04   \\
Jet & 5.54 &  0.60   \\
Lobes & 6.56 &  1.25  \\
\hline
\end{tabular}}
\end{table}
\par The decomposition of the TBCs in terms of OBCs is simple at 330 MHz. The lobe flux
density is everything except the peak flux density of A1, due to the
large beam size compared to the separation ($\sim 3$ times that)
between C0 and C2. The lobe flux density at 1.66~GHz is chosen to be
a sum of: (1) C3+L3, (2) the integral flux density $-$ peak flux density
of C2+L2 and (3) a fraction of the peak of C2+L2. Determining this
fraction is the main unknown in choosing the decomposition of the
TBCs in terms of OBCs, and is chosen to provide the best fit to the
theoretical models. Unless the fraction is at least $\sim 0.5$ in
the lobes, $\alpha$ of the lobes is unrealistically large.
Conversely, full assignment to the lobes produces worse fits overall,
and does not seem consistent with the jet morphology. However, it does
reduce $\alpha$ to 1.05. The two conclusions of relevance are that
there is clearly an excess of flux density at 151~MHz over that
expected from a core-jet system, and the spectral index of the total
flux is 0.77 from 151 MHz to 330 MHz. Secondly, the estimated lobe
flux density and $\alpha$ are virtually identical in our two best
model fits in Table 1. Thusly motivated, using Equations (1)$-$(3),
we estimate that $\overline{Q} = 5.5\pm1.3\times 10^{46}
\rm{ergs~s^{-1}}$.

\section{Accretion Flow Luminosity and Black Hole Mass}
Because of the bright radio core, the optical/UV spectrum might be
contaminated by the tail of the synchrotron spectrum. In fact, the
continuum of the optical spectrum from \citet{smi80} corrected for
Galactic extinction in NED (and the factor of 10 typographical error
on the vertical axis of their spectrum) is consistent with an
extrapolation of the mid-IR power law from \citet{pod15}, as
indicated in Figure~3. Thus, the emission lines are a preferred
method, compared to the continuum, for estimating blazar
$L_{\rm{bol}}$ \citep{cel97}. Using the de-reddened MgII line
strength from \citet{smi80} and the formula from \citet{pun16},
$L_{\mathrm{bol}} \approx 151L(\mathrm{MgII}) \approx 1.71 \pm 0.85
\times 10^{46} \rm{ergs~s^{-1}}$, where the line strength is
$L(\mathrm{MgII}) = 1.14 \pm 0.57 \times 10^{44} \rm{ergs~s^{-1}}$.
Note that the estimate does not include reprocessed radiation in the
infrared from distant molecular clouds. This would be double
counting the thermal accretion emission that is reprocessed at
mid-latitudes \citep{dav11}.

\begin{figure}
\includegraphics[width=70 mm]{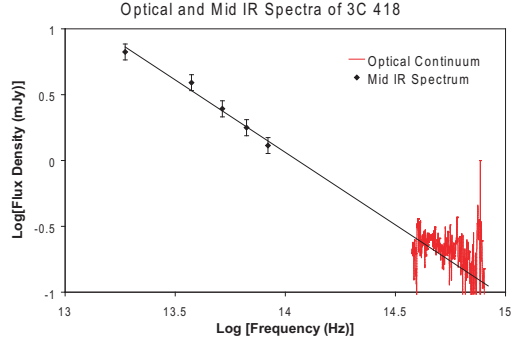}
\caption{The continuum of the optical spectrum is consistent with a
continuation of the mid-IR power law. The data has been corrected for
Galactic extinction.}
\end{figure}

We independently verified the fit to the MgII line in \citet{smi80}, obtaining
similar results including a full width half maximum (FWHM) of $\approx
4400$ km~s$^{-1}$. We insert this value into the virial black hole mass
estimates of \citet{she12} to find
\begin{eqnarray}
&&\log\left(\frac{M_{bh}}{M_{\odot}}\right) =\nonumber\\
&& 3.979 +0.698\log\left(\frac{L(\mathrm{MgII})}{10^{44} \,\rm{erg/s}}\right)
+ 1.382\log\left(\frac{\rm{FWHM}}{\rm{km/s}}\right)\;,
\nonumber\\
&& \frac{M_{bh}}{M_{\odot}} = 1.13 \times 10^{9}\;.
\end{eqnarray}
Alternatively, the formula of \citet{tra12} yields a different
estimate
\begin{eqnarray}
&& \frac{M_{bh}}{M_{\odot}} = \nonumber \\
&& 6.79\times 10^{6} \left(\frac{L(\mathrm{MgII})}{10^{42}
\,\rm{erg/s}}\right)^{0.5}\left(\frac{\rm{FWHM}}{1000
\,\rm{km/s}}\right)^{2},\nonumber\\
&& \frac{M_{bh}}{M_{\odot}} = 6.37 \times 10^{8}\;.
\end{eqnarray}
\section{Discussion}
In this paper, we analyze images at 151~MHz, 330~MHz, 1.66~GHz and
4.86~GHz in order to estimate the diffuse radio lobe flux of the
core-dominated quasar 3C\,418. We estimate a long term time averaged
jet power of $\overline{Q} = 5.5 \pm 1.3 \times 10^{46}
\rm{ergs~s^{-1}}$. {Using the MgII broad emission line,  we
estimated $L_{\mathrm{bol}} \approx 1.71 \pm 0.85 \times 10^{46}
\rm{ergs~s^{-1}}$ and a central super-massive black hole mass of
$M_{bh}/M_{\odot} = 8.9 \pm 2.4 \times 10^{8}$. These results
indicate a very rare state of kinetic dominance for a quasar,
$1.6<\overline{Q}/L_{\rm{bol}}<7.9$ and $0.29
<\overline{Q}/L_{\rm{Edd}}< 0.85$. There are only $\sim 1 - 10$
quasars known to have $\overline{Q}/L_{\rm{bol}}$ this large
\citep{pun07}.

\begin{figure}
\includegraphics[width=72 mm]{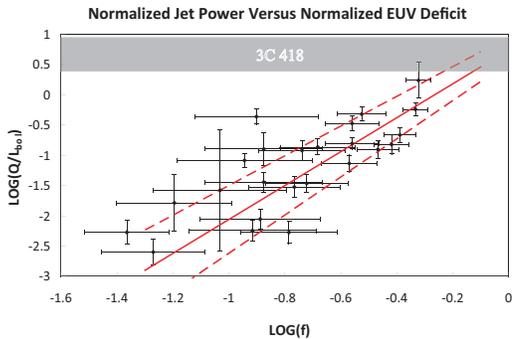}
 \caption{A log-log scatter plot of the fraction of displaced gas, $f$,
 versus $\overline{Q}/L_{\rm{bol}}$ (Punsly 2015). The weighted least squares fit and the standard error are the solid and dashed lines, respectively. The
 range of allowed values of $\overline{Q}/L_{\rm{bol}}$ for 3C\,418 is indicated by the grey band.}
    \end{figure}

The large value of $\overline{Q}/L_{\rm{bol}}$ is  a most extreme
case in the context of the extreme ultraviolet (EUV) deficit of RLQs
relative to RQQs at matched far and near UV luminosity
\citep{zhe97,tel02,pun15}. The stronger the radio jet, the steeper
the spectrum of the EUV continuum, hence the deficit in the EUV for
RLQs. The EUV emission originates from the innermost optically thick
regions of the accretion disk, $<5M_{bh}$ (in geometrized units)
from the inner edge \citep{pun16}. The steep EUV spectrum in consort
with the increased jet power is explained naturally by magnetically
arrested accretion (MAA) in which islands of large scale vertical
magnetic flux create a Poynting flux dominated wind that removes
angular momentum and energy from the innermost portions of accretion
flow \citep{pun15}. These islands are regions of suppressed
turbulence that simultaneously displace the EUV emitting gas,
thereby reducing the EUV emissivity which is created by turbulent
dissipation in the innermost regions of the accretion disk. We
denote the fraction of displaced EUV emitting gas in the innermost
accretion flow by $f$. By comparing to the EUV spectra of RQQs, the
fraction of displaced gas in each RLQ can be estimated. Figure 4 is
a log-log scatter plot from \citet{pun15} of $f$ versus
$\overline{Q}/L_{\rm{bol}}$. The plot was updated with the
additional quasars added in \citet{pun17}, Figure 4. The solid red
line is the weighted least squares fit with uncertainty in both
variables and the dashed lines are the standard error to the fit
\citep{ree89}. Note that 3C\,418 lies in a region that would
indicate $f \LA 1$, an almost complete saturation of the inner
accretion disk by magnetic flux in the MAA model. The fact that the
largest $\overline{Q}/L_{\rm{bol}}$ values implied by observations
are consistent with the physically allowed maximum flux saturation
is strong support for the MAA model of the EUV deficit observed
empirically in RLQs.
\par  In \citet{pun15}, we discussed the
different types of magnetic flux evolution that occur in various
magnetically arrested simulations. The MAA dynamics posited in
\citet{pun15} are based on the simulations in \citet{igu08,pun09} in
which magnetic islands are not extremely short-lived transient
features in the innermost accretion flow. The MAA scenario was shown
to explain the EUV deficit of RLQs, the range of $f$ and the scaling
law in Figure 4. By contrast, other simulations that are described
as ``magnetically arrested" in \citet{ava16} and references therein,
support magnetic islands in the innermost accretion flow, but only
as rare, brief transients and therefore do not explain the EUV
deficit of RLQs. The transient ``prominence" states (a true magnetic
island near the black hole) they describe, would need to be very
common and of longer duration in order to be consistent with
observation. The different simulated behaviors depend on the
physical elements that determine the formation of the magnetic
islands and the time evolution of the magnetic islands in the disk,
reconnection and the diffusion of mass onto and off of the field
lines \citep{igu08,pun15}. The diffusion rate of plasma onto and off
of magnetic field lines and magnetic reconnection rates are not well
known nor well modeled near black holes. These occur in the
simulations as a consequence of numerical diffusion (not a result of
a realistic physical plasma model) in the over-simplified, ideal
magnetohydrodynamic, single fluid models of the physics
\citep{pun15}.  The observations of the EUV deficit can be a
valuable guide for future physical models.
\section*{acknowledgements}
\footnotesize{The National Radio Astronomy Observatory is a facility
of the National Science Foundation operated under cooperative
agreement by Associated Universities, Inc. We thank the staff of the
GMRT who have made these observations
  possible. GMRT is run by the National Centre for Radio Astrophysics
  of the Tata Institute of Fundamental Research.}

\end{document}